\documentclass[
  journal=pasa,
  manuscript=research-paper,
  year=2025,
  volume=42,
]{cup-journal}

\usepackage{listings}
\usepackage{amsmath}
\usepackage[nopatch]{microtype}
\usepackage{booktabs}

\title{Altitude Estimation of Radio Frequency Interference Sources via Interferometric Near Field Corrections}

\author{Jade M. Ducharme}
\affiliation{Department of Physics, Brown University, Providence, RI 02912, USA}
\email[Jade M. Ducharme]{jade\_ducharme@brown.edu}

\author{Jonathan C. Pober}
\affiliation{Department of Physics, Brown University, Providence, RI 02912, USA}

\keywords{instrumentation: interferometers– methods: data analysis– cosmology: observations– dark ages– reionisation– first stars} 

\begin{document}

\begin{abstract}
Radio-frequency interference (RFI) presents a significant obstacle to current radio interferometry experiments aimed at the Epoch of Reionization. RFI contamination is often several orders of magnitude brighter than the astrophysical signals of interest, necessitating highly precise identification and flagging. Although existing RFI flagging tools have achieved some success, the pervasive nature of this contamination leads to the rejection of excessive data volumes. In this work, we present a way to estimate an RFI emitter's altitude using near-field corrections. Being able to obtain the precise location of such an emitter could shift the strategy from merely flagging to subtracting or peeling the RFI, allowing us to preserve a higher fraction of usable data. We conduct a preliminary study using a two-minute observation from the Murchison-Widefield Array (MWA) in which an unknown object briefly crosses the field of view, reflecting RFI signals into the array. By applying near-field corrections that bring the object into focus, we are able to estimate its approximate altitude and speed to be $11.7$ km and $792$ km/h, respectively. This allows us to confidently conclude that the object in question is in fact an airplane. We further validate our technique through the analysis of two additional RFI-containing MWA observations, where we are consistently able to identify airplanes as the source of the interference.
\end{abstract}

\section{Introduction} \label{sec:introduction}

The Epoch of Reionization (EoR) marks a pivotal era in the history of the universe and is characterized by the formation of the first stars and galaxies. Understanding the EoR provides crucial insights into the evolution of large-scale structures and the expansion dynamics of the universe. 21cm cosmology presents itself as a useful tool to probe the EoR by using the hyperfine splitting of neutral hydrogen's 21cm line. Measuring the 21cm signal would effectively allow us to produce three-dimensional maps of the structure of neutral hydrogen during the EoR, which would prove invaluable to our understanding of the dynamics and timeline of reionization. For in-depth reviews of 21cm cosmology and the EoR, the interested reader is referred to \cite{Furlanetto_2006}; \cite{Morales_2010}; and \cite{Liu_2020}.

A direct measurement of the 21cm signal from the EoR has however yet to be made. The faintness of the 21cm signal compared to the overpowering foregrounds and systematics requires an elaborate instrument design and exquisite precision in all ensuing data processing and analysis steps. Current large-scale experiments designed to measure the 21cm signal include the Murchison Widefield Array\endnote{\url{https://www.mwatelescope.org/}} (MWA; \citealt{mwa1}, \citealt{mwa2}), the Hydrogen Epoch of Reionization Array\endnote{\url{https://reionization.org/}} (HERA; \citealt{DeBoer_2017}), and the LOw Frequency ARray\endnote{\url{https://www.astron.nl/telescopes/lofar/}} (LOFAR; \citealt{lofar}).

In addition to accounting for the instrument and all its associated systematics, the analysis pipeline must also account for sporadic and unpredictable radio-frequency interference (RFI) signals. These signals can have a variety of origins and have for example been traced back to broadband digital television (DTV) and digital audio broadcasting (DAB) signals \citep{Wilensky_2019, Wilensky_2020}, direct emissions from satellites \citep{DiVruno_2023, Grigg_2023}, and reflections off of satellites \citep{prabu2023}.

Traditionally, RFI signals have been dealt with on a per-observation basis by identifying the contaminated time and frequency channels and flagging these, effectively removing them from further analysis. The \texttt{AOFlagger} \citep{AOFlagger} and \texttt{SSINS} \citep{Wilensky_2019} tools in particular have been used extensively with MWA data and have proven effective in flagging regions of RFI ranging from bright to faint. The problem arises when ultrafaint RFI, undetectable by any of the above tools, remains unseen in the data, contaminating our data on a scale that is larger than that of the 21cm signal itself and thus obscuring it. Recent studies have in fact shown that, on the basis of RFI contamination alone, over 50\% of all available MWA data may need to be discarded \citep{Wilensky_2023}.

In this paper, we present an approach to localizing sources of RFI in MWA observations using near-field corrections. Our method, which builds upon the work presented in \cite{prabu2023}, combines far-field phasing, near-field corrections, and beamforming in order to maximally cohere (or ``focus'') the curved near-field signals from the RFI source as they are received by the MWA antennas. Accurately determining the precise three-dimensional location of RFI emitters could enable their targeted removal as opposed to simple flagging, thereby preserving a greater proportion of the usable data.

This research does not provide a detailed quantitative assessment of the RFI source altitude obtained through our methods. Instead, we demonstrate that the altitude can be identified and reasonably constrained using basic statistical techniques. Future work will focus on a more in-depth quantitative analysis, as a precise understanding of these statistics will be essential for determining the limits of RFI subtraction or peeling in the context of EoR research. For instance, a recent study demonstrated that an effective RFI subtraction strategy can reduce data loss to as little as 1\%, a significant improvement over the aforementioned potential 50\% lost through flagging alone \citep{Finlay_2023}. However, since that study did not specifically focus on EoR research, it remains uncertain whether this level of precision will be sufficient for EoR applications.

This paper is organized as follows: in Section \ref{sec:core_concepts}, the relevant core concepts are presented. In Section \ref{sec:methods}, we detail our methodology. In Section \ref{sec:results}, we present our results, and produce concluding remarks in Section \ref{sec:conclusion}.

\section{Core Concepts} \label{sec:core_concepts}

This section introduces the essential concepts and definitions necessary for understanding the techniques and results discussed in the following sections.

\subsection{Far-Field Phasing}\label{phasing}

In this work, we distinguish between two interferometric phasing techniques by referring to the commonly known ``phasing'' as ``far-field phasing'' to clearly differentiate it from the ``near-field phasing'' technique, also discussed herein.

Far-field phasing refers to the adjustment of the phase of interferometric visibilities in order to cohere signals from a source located in the far field of the instrument. It requires a precise knowledge of the location of the source in the local reference frame in order to make the necessary adjustments. These adjustments can be made on the physical instrument itself --- e.g. by adding or removing cable segments to account for the relative light travel time between the various antenna elements of the array --- or, alternatively, they can be made computationally after the data has been collected, which is the method used in this work. Specifically, we use the interferometric Python package  \texttt{pyuvdata}\endnote{\url{https://github.com/RadioAstronomySoftwareGroup/pyuvdata}} \citep{pyuvdata} to perform all far-field phasing corrections to the data used in this work.

Interferometric arrays perform correlations between antennas under the assumption that the sources are infinitely far away such that the signals hit the receivers as plane waves \citep{Thompson2001}. This assumption simplifies far-field phasing calculations by negating the need to account for the curvature of light wavefronts, which is negligible for astronomical sources.

Ionospheric distortions can affect the far-field plane wave assumption by introducing small shifts in source positions. These distortions are generally mild, but during periods of severe ionospheric conditions, higher-order effects may occur, making the data unreliable. In such cases, the affected data is typically discarded.

\subsection{Beamforming}

In the context of radio interferometry, the term ``beamforming'' can refer to various processes. For clarity, within this paper, beamforming specifically refers to the procedure of averaging all visibilities to concentrate the array beam into a single ``pixel'' in image space, thereby isolating the intensity of the sources contained within that pixel. When beamforming, the contributions from other pixels cancel each other out. This is because far-field phasing sets the phase of emissions from the phase center to zero for every baseline, maintaining coherence when averaged; for sources not at the phase center, the phase varies across baselines, resulting in their contributions averaging down to significantly smaller values than those at the phase center. However, beamforming sidelobes can introduce bias, which may be substantial relative to faint signals, such as those from the EoR. Although this could be problematic if we were aiming to derive an exact numerical value for the beamformed intensity, our focus is only on the maximum beamformed intensity after iterating through parameters that are assumed not to affect the sidelobe structure (see Section \ref{sec:estimate}).

Beamforming therefore offers a way to estimate the intensity of any radio source simply by applying a far-field phase offset to the dataset using the coordinates of this source, and averaging the resulting visibilities together.

\subsection{Near-field corrections}

 While the far-field assumption discussed in Section \ref{phasing} may hold for stars, galaxies, and other astronomical radio objects, it breaks down for sources of RFI which are usually located much closer to the array. Light emanating from these so-called near-field objects hits the array as a spherical wave whose curvature can no longer be ignored.

 In the far field, where all wavefronts are coplanar, the phase difference of the wave at each antenna is determined solely by its angular position in the sky. However, in the near field, where wavefronts are curved, there is an additional phase difference dependent on the source distance. Since standard imaging pipelines make the far-field assumption, all the phase differences are interpreted as coming purely from source position. This leads to different baselines putting the source at different angular positions during image reconstruction, producing blurry, out-of-focus images.

It is however possible to apply near-field corrections to near-field objects in order to bring them into focus \citep{marr2015fundamentals}. The near-field corrections work by first applying a far-field phase offset to the data using the right ascension and declination of the object. This is done in order to ``center'' the data, allowing us to produce near-field corrections as additional fine-grained deviations to this coarse phasing. The near-field corrections themselves are calculated by assuming a radial focal distance, $f$, between the array and the object, and calculating the geometrical corrections resulting from the spherical symmetry of the emitted light. The following calculations are discussed in detail in \cite{prabu2023}, but we provide a brief summary here.

First, the object's coordinates in the Topocentric Cartesian Coordinate (TCC) system that is centered on the array are calculated using Equation \ref{focal_dist}.

\begin{equation}\label{focal_dist}
    \begin{aligned}
        & f_x=f_{\text {dist }} \times \cos (\phi) \times \sin (\theta), \\
        & f_y=f_{\text {dist }} \times \sin (\phi), \\
        & f_z=f_{\text {dist }} \times \cos (\phi) \times \cos (\theta) ,
    \end{aligned}
\end{equation}
where $f_{\text{dist}}$ corresponds to the assumed focal distance; $f_i$ to the projected focal distance along the $i$-th axis; and $\phi$ and $\theta$ to the azimuthal and polar angles of the object in the TCC frame.

The delay ($w$-term) for each baseline (corresponding to the correlation between antennas $i$ and $j$) is then obtained using Equation \ref{w_term}.

\begin{equation}\label{w_term}
    \begin{aligned}
    & r_i=\sqrt{\left(f_x-X_i\right)^2+\left(f_y-Y_i\right)^2+\left(f_z-Z_i\right)^2}, \\
    & r_j=\sqrt{\left(f_x-X_j\right)^2+\left(f_y-Y_j\right)^2+\left(f_z-Z_j\right)^2}, \\
    & w_{\text {near-field }, i, j}=r_j-r_i ,
    \end{aligned}
\end{equation}
where $(X_i,Y_i,Z_i)$ and $(X_j,Y_j,Z_j)$ correspond to coordinates in the TCC frame of antennas $i$ and $j$.

Finally, the near-field phase correction is calculated using Equation \ref{correction}.

\begin{equation}\label{correction}
    \Delta \phi_{i, j}=\exp ^{i 2 \pi \frac{\Delta w_{i, j}}{\lambda}},
\end{equation}
where $\Delta w_{i,j} = w_{\text{near-field},i,j} - w_{\text {far-field }, i, j}$ and $\lambda$ corresponds to the observed wavelength.

This correction term is dependent on the selected baseline, frequency and time, and must therefore be calculated for each individual visibility.

\cite{prabu2023} showed that the optimal focal distance results in a maximally focused image with the highest SNR. They show that it is possible to estimate an object's altitude by iterating over a range of focal distances, imaging the corrected visibilities, and calculating the SNR, recording the focal distance for which the SNR is maximal.

\subsection{Beamforming and near-field corrections}

In this work, we combine the concepts of beamforming and near-field corrections in order to skip the iterative imaging step, thereby saving time and computing resources. Indeed, we will show that by iterating through a range of focal distances and beamforming at each step,  the resulting intensity is maximized at the focal distance corresponding to the object's location. By skipping the imaging step at each focal distance, our method is orders of magnitude faster while retaining a similar accuracy.

Although our method does not include a secondary iterative step to refine the focal distance estimate after the initial maximization, this could be considered for future optimization. However, as discussed in Section \ref{sec:results}, precision is not the primary limitation on accuracy in this work. Factors such as the MWA antenna configuration and time resolution play a more significant role in determining the accuracy of the focal distance. As a result, an additional optimization step would likely provide diminishing returns.

Our method still relies on obtaining accurate right ascension and declination coordinates of the near-field object for each time-step. We describe how we obtain these in detail in the following section.

\section{Methods/Data Analysis} \label{sec:methods}

\begin{table*}[ht]
    \centering
    \begin{tabular}{|c|c|c|c|}
        \hline
       \textbf{OBSID}  & \textbf{1061313128} & \textbf{1092761680} & \textbf{1252945816} \\
       \hline
       Observation date  & August 2013 & August 2014 & September 2019 \\
       \hline
       Antenna configuration & Phase I & Phase I & Phase II (Compact)\\
       \hline
       Time resolution (s) & $0.5$ & $2.0$ & $2.0$ \\
       \hline
       Duration (s) & $112$ & $112$ & $120$ \\
       \hline
       Number of time-steps containing RFI & $59$ & $15$ & $19$ \\
       \hline
       Frequencies containing RFI (MHz) & $181.5-187.5$ & $181.5-187.5$ & $181.5-187.5$ \\
       \hline\hline
       Average object altitude (km) & $11.7 \pm 0.1$ & $11.73 \pm 0.05$ & $13.9 \pm 0.9$ \\
       \hline 
       Average object speed (km/h) & $792 \pm 1$ & $1050 \pm 20$ & $1360 \pm 30$ \\
       \hline
    \end{tabular}
    \caption{Detailed information for each MWA observation included in our study. The rows above the double-line break (e.g., OBSID, observation date, antenna configuration, etc.) describe the observational characteristics. The rows below the double-line break provide the RFI emitter's characteristics, determined using the technique presented in this work. Note that the uncertainties recorded here are obtained using the standard error on the mean. While they provide a good measure of the technique's precision, they do not account for potential inaccuracies.}
    \label{tab:info}
\end{table*}

sThe data used in our preliminary study is a Phase I MWA observation downloaded from the MWA's All-Sky Virtual Observatory (ASVO)\endnote{\url{https://asvo.mwatelescope.org/}}. Its ASVO observation ID (OBSID) is 1061313128, but for simplicity we will refer to it as the target observation throughout this paper. It is a two-minute observation targeting the EoR0 field, taken on 2013-08-23 and operating over 768 frequency channels between 167 and 198 MHz. This observation was selected because it was featured in \cite{Wilensky_2019} and therefore known to contain RFI. Through ASVO, the target was converted to the CASA Measurement Set format \citep{AOFlagger} with a time resolution of 0.5 seconds and a frequency resolution of 40 kHz. The option to apply a basic calibration solution was selected \citep{calib}. Additionally, a second round of direction-independent calibration was performed using \texttt{MWA-Hyperdrive}\endnote{\url{https://github.com/MWATelescope/mwa_hyperdrive}} in order to obtain data that is sufficiently well calibrated for imaging purposes. The calibration was performed using a source list\endnote{\url{https://github.com/JLBLine/srclists?tab=readme-ov-file}} created using the LoBES catalog \citep{Lynch_2021}. The 8000 brightest sources within 120 degrees of the phase center are used in calibration, and bad tiles, found using the quality assurance method presented in \cite{Nunhokee2024}, were flagged in order to produce the optimal calibration solutions. \texttt{MWA-Hyperdrive} was also used to subtract 8000 background sources from the visibilities using the same source list. This was done in order to increase the signal-to-noise ratio in image space and make the RFI signal stand out more brightly. Although this sky subtraction is not strictly required for the analysis, we found anecdotally that the imaging software's component fitting algorithm—described in the next paragraph—performs more effectively after this step.

\begin{figure}[hbt!]
\centering
\includegraphics[width=\linewidth]{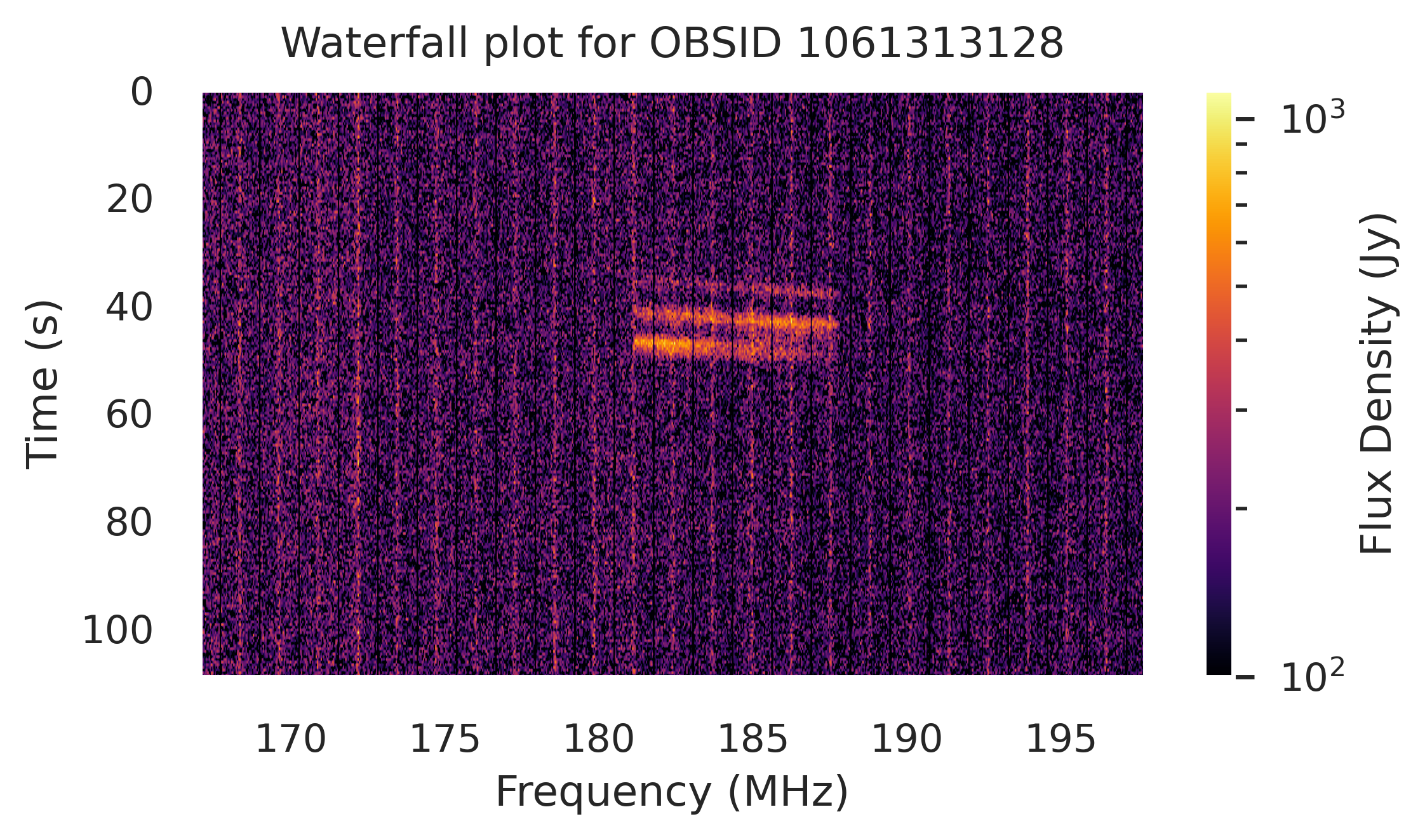}
\caption{Single-baseline, two-dimensional time vs. frequency waterfall plot for our target observation. We can visually identify the RFI region contained between frequencies of 181.5 MHz to 187.5 MHz, and times between 35 seconds and 50 seconds; however, extending the selected time range has revealed that the RFI is still visible in image-space for time-steps that appear contamination-free in the waterfall plot. This is made obvious in the movie found in the accompanying material to this article, where the the object is successfully imaged and tracked over 59 time-steps.}
\label{waterfall}
\end{figure}

Our altitude estimation method requires a good estimate of the RFI emitter's position in the sky; in other words, its right ascension and declination coordinates. We obtain these through imaging via \texttt{WSClean}\endnote{\url{https://gitlab.com/aroffringa/wsclean/}} \citep{offringa-wsclean-2014}. Specifically, we use \texttt{WSClean}'s component fitting method, enabled via the \texttt{-save-source-list} flag, which produces estimates for the sky coordinates of all sources present in the data. However, imaging the complete data set is not only computationally expensive and time-consuming, but it also often fails to identify the RFI emitter. This is because the emitter only produces a signal during specific time-steps and in specific frequency channels, and imaging over the whole band tends to wash out this signal. It is therefore necessary, as a first step, to produce a two-dimensional time vs. frequency waterfall plot to help localize the RFI. A single-baseline waterfall plot example for our target observation (after calibration and subtraction) is shown in Figure \ref{waterfall}.

In the figure, we can visually identify the contaminated time and frequency channels where the flux density is significantly higher. A typical analysis pipeline would flag and remove these channels, effectively eliminating the brightest RFI but potentially missing RFI emissions that fall below the noise level. For instance, we were able to image the RFI emitter in our target observation even during time-steps that did not show elevated flux density in the waterfall plot. Whether flagging fully removes RFI is debatable, but it certainly also discards valuable underlying sky signals. To avoid this loss, the focus of this work is on laying the groundwork for methods that directly subtract the contamination from the visibilities, allowing these channels to remain usable and propagate through the rest of the analysis.

\begin{figure*}
    \centering
    \includegraphics[width=\linewidth]{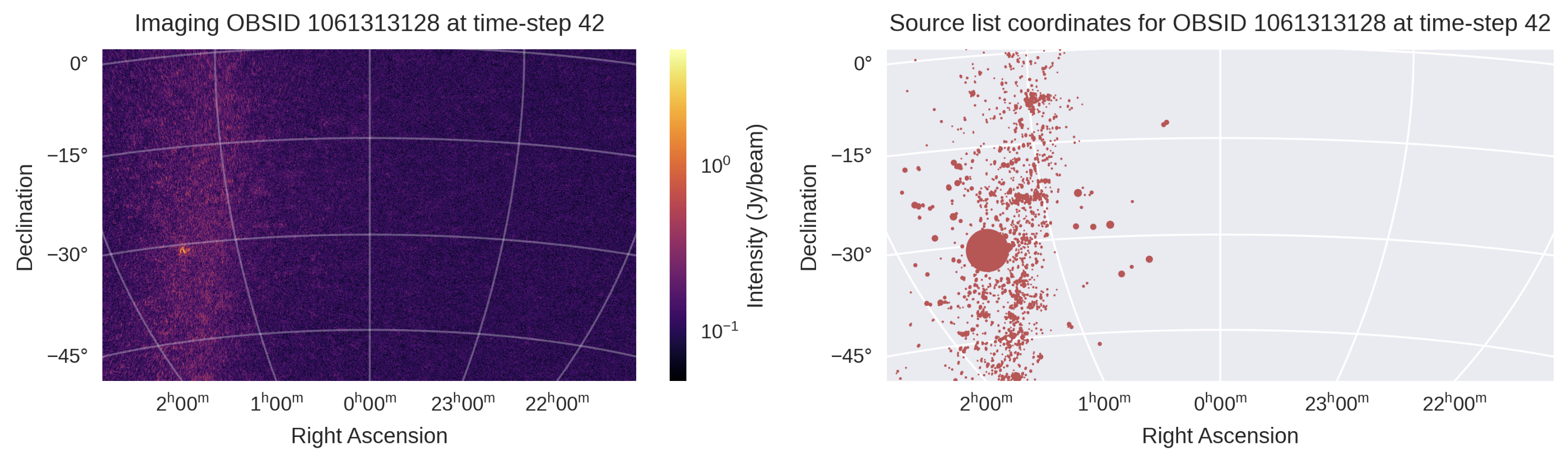}
    \caption{(Left) Image output by \texttt{WSClean} for our target observation at an arbitrarily chosen time-step. (Right) Source list coordinates returned by \texttt{WSClean} for the same time-step. The dot sizes are proportional to the intensity of the source as measured by \texttt{WSClean}. We note that the RFI emitter is significantly brighter than all other recorded sources, and remains so for all time-steps of interest not pictured here.}
    \label{fig:concat}
\end{figure*}

Once the time and frequency bands containing RFI have been identified, the original dataset is then sampled down to the frequency channels and time-steps containing the signal. Each individual time-step is then imaged using \texttt{WSClean}'s multi-scale, multi-frequency, and auto-masking algorithms \citep{offringa-wsclean-2017} in order to obtain accurate source component lists containing the source's right ascension and declination coordinates. An example of the image and source list output by \texttt{WSClean} is shown in Figure \ref{fig:concat} for a randomly selected time-step.

As can be seen in the figure, it is very straight-forward to identify the RFI emitter's coordinates in the source list given that it is always the brightest source at all time-steps. Having then obtained these coordinates, we move on to implementing the altitude estimator.

\subsection{Estimating the altitude via near-field corrections} \label{sec:estimate}

The near-field corrections for one single time-step are obtained via the following steps:

\begin{itemize}
    \item Apply far-field phasing to the current RA and DEC of the object;
    \item Calculate and apply near-field 
        corrections for the desired focal distance $f$ via Equation \ref{correction};
    \item Average all the visibilities together (beamform) -- this value corresponds to the intensity of the RFI emitter at the focal distance $f$.
\end{itemize}

\begin{figure*}
    \centering
    \includegraphics[width=1.0\linewidth]{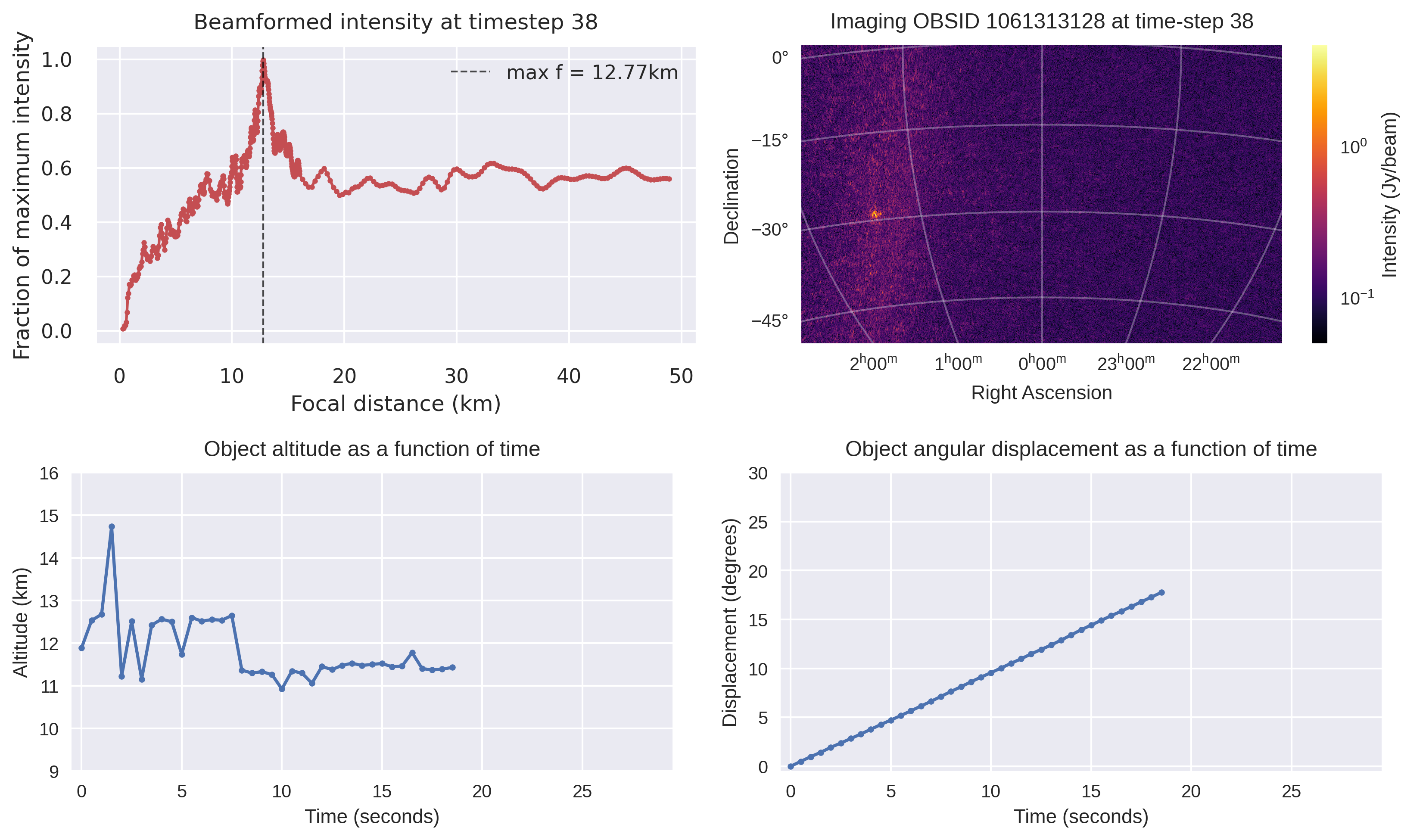}
    \caption{(Top left) The average (beamformed) visibility for a range of different focal distances. The maximum intensity occurs at the optimal focal distance corresponding to our best guess to the airplane's actual location. (Top right) \texttt{WSClean}'s image output, where the airplane can clearly be seen. (Bottom left) The measured altitude as a function of time, where $t=0$ corresponds to the first time-step of interest for this observation. (Bottom right) The airplane's angular displacement as a function of time. The angular displacement is calculated using the Euclidean distance between the angular coordinates of the object at time $t$ vs. its angular coordinates at time $t=0$. A movie where each frame corresponds to a different time-step is available in the supplementary material accompanying this article.}
    \label{fig:total}
\end{figure*}

These steps are repeated for a range of focal distances, creating a list. This list is plotted in the top-left panel of Figure \ref{fig:total}. We note that the maximum intensity corresponds to our best guess for the actual focal distance to the airplane.

These steps are further repeated for every time-step of interest, allowing us to plot the airplane's estimated altitude above sea level as a function of time, shown in the bottom-left panel of Figure \ref{fig:total}. The altitude is obtained by first calculating, via the \texttt{astropy} Python package, the vertical projection of the slant focal distance between the airplane and the MWA. Adding the MWA's own altitude to this value gives us the airplane's altitude above sea level.

The four panels summarizing our results for the thirty-eighth time-step are shown in Figure \ref{fig:total}. This time-step was chosen since it clearly shows the bright RFI emitter in the top-right panel. This object is fainter during earlier and later time-steps. The movie featuring all 59 time-steps is provided as additional material to this article.

The comprehensive analysis, including the initial time-step and frequency channel selection, imaging at each time-step, and subsequent beamforming at each time-step and focal distance, can require a few hours on a 12-core system such as the one we employed. The RAM usage should not exceed the observation's file size, which for our target observation is 64 Gb. The most significant bottlenecks occur during the imaging and iterative beamforming steps. To improve efficiency, further work is needed to parallelize these processes.

\section{Results} \label{sec:results}

Using beamforming in combination with the near-field corrections presented in \cite{prabu2023} allows us to efficiently obtain an estimate for the altitude of a near-field radio-emitting object such as an airplane or a satellite. For our target observation, this technique allowed us to estimate the object's average altitude to approximately $11.7 \pm 0.1$ km, suggesting its likely classification as an airplane since a plane's cruising altitude can vary between 9.4 and 11.6 km \citep{SFORZA201447}. To further support this hypothesis, we used the angular displacement as a function of time in combination with the measured altitude to calculate its speed, which we find to be $792 \pm 1$ km/h, also consistent with an airplane's cruising speed.

The uncertainties presented here are simply calculated using the standard error on the mean, as in Equation \ref{eq:error}.

\begin{equation} \label{eq:error}
    \textrm{SE} = \frac{\sigma}{\sqrt{n}},
\end{equation}

where $\sigma$ is the standard deviation across all samples, and $n$ is the sample size. In our case, the standard deviation is calculated across time-steps (i.e. across the data points plotted in the bottom-left panel of Figure \ref{fig:total}), and the sample size corresponds to the number of time-steps. We do not include errors associated with the coordinates found by  \texttt{WSClean} (bottom-right panel of Figure \ref{fig:total}) or with the beamforming process itself (top-left panel of Figure \ref{fig:total}). As such, the errors reported here provide a reliable measure of our technique's precision, but they do not account for potential inaccuracies.

Attempts to track down the exact flight captured in the target observation have unfortunately proved fruitless. Numerous airspace APIs, such as \texttt{AeroAPI}\endnote{\url{https://www.flightaware.com/commercial/aeroapi/}} or \texttt{FlightAPI}\endnote{\url{https://www.flightapi.io/}}, are available online for the purpose of identifying specific aircraft using their altitude, latitude, longitude, and many other features. However, we have not found one whose historical data extends back to 2013, when the target observation was made. Additionally, there is no guarantee that the aircraft sighted in the observation is trackable using a public API, as it could be a private plane not listed in the public airspace database.

\subsection{Additional Observations}

In order to verify the effectiveness of our technique, we also applied it to two other MWA observations, selected after manual inspection of their waterfall plots revealed a similar RFI structure to the one presented in Figure \ref{waterfall}. For all observations, including the target observation discussed thus far, we provide the OBSID, target field, time resolution, and other relevant details in Table \ref{tab:info}. Based on the ranges of altitudes and speeds obtained for all these observations, also provided in Table \ref{tab:info}, we note that, as was the case for the target observation discussed above, these all most likely correspond to airplane reflection events.

\begin{figure*}[hbt!]
    \centering
    \includegraphics[width=\textwidth]{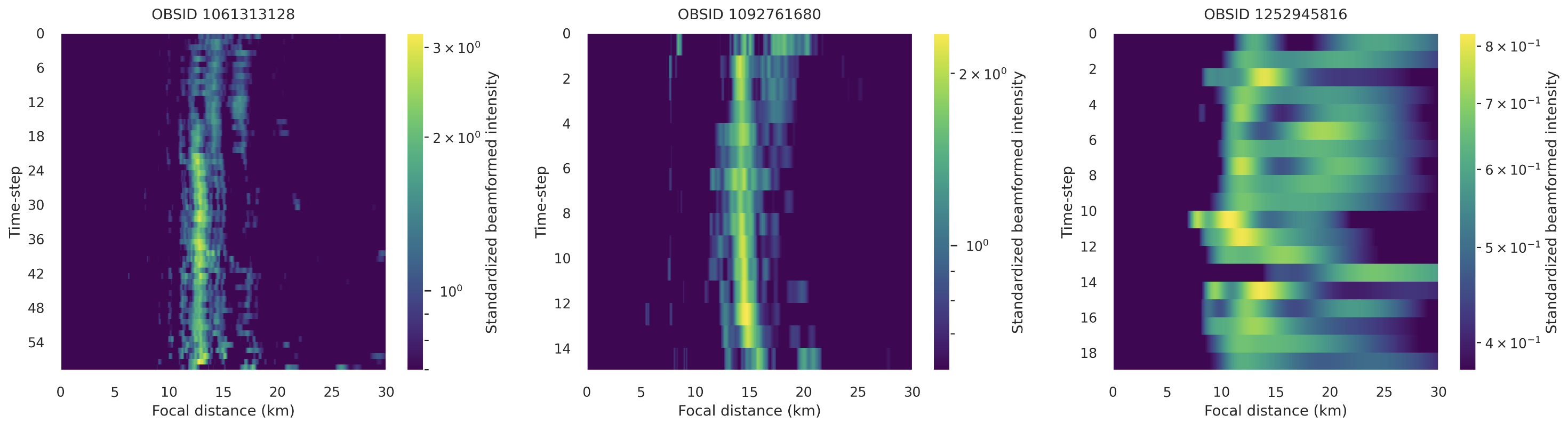}
    \caption{Standardized beamformed intensities per time-step and focal distance for all three observations. Note that all plots are on a logarithmic color scale. Lower-end values were truncated to help highlight prominent features and provide a smooth color scale transition. Higher standardized beamformed intensity values correspond to higher confidence that the RFI emitter is found at the corresponding focal distance. Details concerning the standardization and truncation process are provided in Section \ref{sec:results}.}
    \label{fig:final_beamformed_waterfall}
\end{figure*}

Figure \ref{fig:final_beamformed_waterfall} presents a summarized view of our results. It shows the standardized beamformed intensity per time-step and focal distance for all three observations. The standardization process for the beamformed intensity at each time-step is performed using the following equation:

\begin{equation}
    I_s = \frac{I - \hat{I}}{\sigma_I},
\end{equation}
where $I$ corresponds to the raw beamformed intensity at a specific time-step, $\hat{I}$ to the mean beamformed intensity at that time-step, and $\sigma_I$ to the standard deviation at that time-step. This standardization, repeated for each time-step, allows us to control the dynamic range of the intensities, facilitating their visualization.

To further enhance visualization, we remove lower-end values, allowing us to present these plots on a logarithmic scale.  This highlights the more prominent features of the data. The data truncation for each observation is achieved by defining a threshold value, below which all data points are set to the threshold. This ensures a smooth color scale transition. The chosen threshold values for OBSIDs 1061313128, 1092761680, and 1252945816 are 2.5, 2, and 0.45 standard deviations below the maximum standardized beamformed intensity, respectively. These values were determined through trial-and-error to best highlight the significant features.

From Figure \ref{fig:final_beamformed_waterfall}, it is evident that the third observation from the left performs significantly worse than the first two. The maximum beamformed intensity recorded is much lower, and the intensity spread is larger, indicating reduced overall confidence in the results for this observation.

\section{Concluding Remarks} \label{sec:conclusion}

Using far-and-near-field corrections in combination with beamforming allows us to make precise estimates of an RFI emitting object's altitude. Being able to accurately localize RFI emitters will prove invaluable as more and more researchers focus on extracting or ``peeling'' RFI as opposed to simply flagging it. This extraction strategy has the advantage of allowing us to preserve a larger fraction of all available data, maximizing the likelihood of a 21cm signal detection.

We find that although our technique performs well overall, there are notable differences in performance across observations. Indeed, as can be noted from the error bars in Table \ref{tab:info}, our altitude and speed estimates are significantly less precise for the third observation from the left compared to the first two. The spread in Figure \ref{fig:final_beamformed_waterfall} is also much more pronounced for this observation.

We hypothesize that this is due to the difference in the array configuration for this observation. Indeed, the first two observations are in the MWA Phase I configuration, featuring several long baselines, whereas the third observation is in the Phase II (compact) configuration mostly comprised of short baselines. For shorter baselines, for which the light travel time between antennas is shorter, the far-field assumption holds at a closer distance compared to longer baselines. This was also noted in \cite{Prabu_2022} in the context of low-earth orbit satellites, which appear in the far field of the MWA for the Phase II (compact) configuration. It is therefore understandable that applying near-field corrections to an object seen in the far field of the intrument does not prove very effective.

There is an additional source of variance coming from the difference in time resolutions (0.5 seconds for the first observation, and 2.0 seconds for the other two). As shown in the supplementary material accompanying this article, the images produced from the 2.0-second resolution data feature smeared and oblong RFI emitters. This occurs because the airplanes move too fast for the given time resolution. This makes obtaining high-precision coordinates challenging, as the airplane appears as a streak over the 2-second interval and cannot be approximated as a point source with a single set of coordinates.

This paper represents the first definitive detection and localization of an airplane in MWA data. Detecting and localizing airplanes is critical for future operations, given the high frequency of flights in the Murchison Radio-astronomy Observatory (MRO) region. Indeed, a recent study conducted in the MRO has found that aircraft are present above the horizon line at least 13\% of the time, establishing a lower limit on the data potentially lost to reflected RFI from aircraft \citep{Tingay_2020}.

Although a detailed quantitative analysis is beyond the scope of this work, the constraints derived from simple statistical methods are promising and indicate that further research, particularly on subtracting and peeling RFI, is a valuable pursuit. Future studies will require a more rigorous quantitative approach to determine whether RFI can indeed be subtracted to a level below the EoR signal.

A large drawback of using the altitude estimation method presented herein is the need to obtain the object's right ascension and declination sky coordinates in order to perform the far-field phasing. This poses a great problem because, as we noted in Section \ref{sec:introduction}, even RFI that is too faint to be seen using current flagging algorithms (and, by extension, current imaging software) is still bright enough to overpower the 21cm signal. Future work will therefore focus on generalizing this technique to faint RFI signals whose origin is not a priori known or easily obtainable.

\begin{acknowledgement}

This scientific work uses data obtained from Inyarrimanha Ilgari Bundara / the Murchison Radio-astronomy Observatory. We acknowledge the Wajarri Yamaji People as the Traditional Owners and native title holders of the Observatory site. Establishment of CSIRO's Murchison Radio-astronomy Observatory is an initiative of the Australian Government, with support from the Government of Western Australia and the Science and Industry Endowment Fund. Support for the operation of the MWA is provided by the Australian Government (NCRIS), under a contract to Curtin University administered by Astronomy Australia Limited. This work was supported by resources provided by the Pawsey Supercomputing Research Centre with funding from the Australian Government and the Government of Western Australia.

We would furthermore like to thank Steve Prabu for providing guidance and clarifications to the near-field phasing technique; Steven Tingay for his helpful suggestions on the first draft of this paper; Michael Wilensky for providing all OBSIDs used in this work; Dev Null for their assistance with \texttt{MWA-Hyperdrive} and \texttt{WSClean}; and Miguel Morales for his valuable suggestions during the biweekly EoR Imaging telecons, which led to the creation of Figure \ref{fig:final_beamformed_waterfall}.

\end{acknowledgement}

\paragraph{Funding Statement}

This research was supported by a grant from the US National Science Foundation (award ID 2228989) which includes a Graduate Research Supplement from the Spectrum Innovation Initiative.  

\paragraph{Competing Interests}

None

\paragraph{Data Availability Statement}

The MWA data used in this work are available to download
at ASVO, https://asvo.mwatelescope.org/

\printendnotes

\bibliography{main}




\end{document}